\documentclass[letterpaper]{article} 
\usepackage[submission]{aaai25}  
\usepackage{times}  
\usepackage{helvet}  
\usepackage{courier}  
\usepackage[hyphens]{url}  
\usepackage{graphicx} 
\usepackage{natbib}  
\usepackage{caption} 
\usepackage{algorithm}
\usepackage{algorithmic}
\usepackage{amsmath}

\usepackage{newfloat}
\usepackage{listings}
\DeclareCaptionStyle{ruled}{labelfont=normalfont,labelsep=colon,strut=off} 
\floatstyle{ruled}
\newfloat{listing}{tb}{lst}{}
\floatname{listing}{Listing}
\title{Releasing the Parameter Latency of Neural Representation for High-Efficiency Video Compression}
\author {
    Gai Zhang$^1$,
    Xinfeng Zhang$^1$,
    Lv Tang$^1$,
    Yue Li$^2$,
    Kai Zhang$^2$,
    Li Zhang$^2$
}
\affiliations {
    $^1$University of Chinese Academy of Sciences\\
    $^2$Bytedance Inc.\\
    zhanggai16@mails.ucas.ac.cn, xfzhang@ucas.ac.cn, luckybird1994@gmail.com, yue.li@bytedance.com, zhangkai.video@bytedance.com, lizhang.idm@bytedance.com
}

\usepackage{bibentry}

\begin{document}

\maketitle

\begin{abstract}
For decades, video compression technology has been a prominent research area. Traditional hybrid video compression framework and end-to-end frameworks continue to explore various intra- and inter-frame reference and prediction strategies based on discrete transforms and deep learning techniques. However, the emerging implicit neural representation (INR) technique models entire videos as basic units, automatically capturing intra-frame and inter-frame correlations and obtaining promising performance. INR uses a compact neural network to store video information in network parameters, effectively eliminating spatial and temporal redundancy in the original video. However, in this paper, our exploration and verification reveal that current INR video compression methods do not fully exploit their potential to preserve information. We investigate the potential of enhancing network parameter storage through parameter reuse. By deepening the network, we designed a feasible INR parameter reuse scheme to further improve compression performance. Extensive experimental results show that our method significantly enhances the rate-distortion performance of INR video compression.
\end{abstract}

\section{Introduction}
With the rise of various User Generated Content (UGC) platforms, the rapid increase of video content on the Internet has led to an exponential growth in video data. Therefore, efficient video compression is crucial for storage and transmission. In the past few decades, video compression technology has made significant progress, resulting in complex codec standards like H.264/AVC \cite{wiegand2003overview}, H.265/HEVC \cite{sullivan2012overview}, and H.266/VVC \cite{bross2021overview}. These traditional techniques rely on methods such as discrete transformation, quantization, entropy coding, and filtering to compress video data. However, due to the intricate design and complexity of each module, traditional video codecs fail to achieve end-to-end rate-distortion optimization, limiting their potential for optimal performance. As video consumption continues to grow, the demand for more advanced and efficient video compression methods is becoming increasingly important.

Breakthroughs in deep learning have revolutionized video compression technology, opening up new possibilities for efficient video compression. End-to-end neural video codecs have achieved notable success by learning compression strategies directly from data in a fully differentiable manner \cite{lu2019dvc,li2021deep}. These neural codecs optimize the encoder, decoder, and entropy model together, resulting in superior rate-distortion performance compared to traditional video compression standards. Despite these advances, training such large-scale neural networks demands substantial amounts of data and computing power, which presents significant challenges for practical deployment, especially on devices with limited resources. Nevertheless, the potential of neural video codecs continues to drive innovation in this area.

The emergence of implicit neural representation (INR) offers a promising new direction for neural video codecs. INR leverages neural networks to represent data compactly and efficiently. INR-based video compression trains a simple neural network to fit the target video and then compresses the network parameters to achieve video compression. This approach significantly reduces model complexity and computational demands at the decoder compared to traditional end-to-end neural codecs. Recent advancements in INR-based video compression have brought rapid improvements in rate-distortion performance. However, these methods still fall short compared to the performance of traditional video codecs and end-to-end neural codecs. \textbf{“Has the information storage potential of INR network parameters been fully utilized?”} Addressing this question is crucial for assessing the effectiveness and practicality of INR-based video compression.

To maximize compression performance, the goal is to retain as much of the original video information as possible while using the least amount of data, which should be encoded. For INR-based video compression, we need to determine whether existing INR network parameters can fully utilize their representation capabilities to maximize the quality of retained videos at a given bit rate. The current INR structure is simple, usually consisting of basic modules (CNN or MLP) stacked together. According to the development history of large neural networks, introducing complex mechanisms such as attention can enable neural networks with the same parameter level to achieve better performance. Therefore, current INR-based video compression methods may not fully unleash the potential of available parameter space, leaving room for further optimization.

Fully unleashing the information storage capability of INR network parameters remains an open problem. In this work, we validate the potential of current INR video compression methods to preserve more video information through a parameter reuse mechanism. We conducted extensive exploration on the parameter reuse mechanism of INR. By strategically reusing and recombining INR parameters, we achieve more expressive and informative video representations, significantly improving compression efficiency. Specifically, we explore a series of parameter reuse schemes, including weight cascading and network deepening, to determine the optimal configuration that maximizes video quality at a given bit rate. Our extensive experiments on various video datasets showed that the proposed method has superior performance compared to exisiting INR-basd video compression methods.

The main contributions of this paper are:
\begin{itemize}
    \item We validate the network parameters of current INR-based video compression methods still have the potential to save more video information through parameter reuse.
    \item We propose a suite of effective parameter reuse techniques to unleash the potential of neural representations for efficient video compression.
    \item We report significant rate-distortion performance improvements on multiple video datasets, demonstrating the ability of our method to enhance the performance of INR video compression methods.
\end{itemize}

\section{Related Works}

\subsection{End-to-End Neural Video Coding}
With continuous breakthroughs in deep learning for computer vision tasks, neural video codecs advance rapidly. DVC \cite{lu2019dvc} marks a significant milestone, achieving end-to-end neural video codec rate-distortion optimization for the first time while still using the traditional residual coding framework. Numerous neural video codecs \cite{djelouah2019neural,hu2020improving,lin2020m,yang2020learning,agustsson2020scale,hu2021fvc,rippel2021elf,hu2022coarse,chen2022lsvc} follow this residual coding approach. For example, Djelouah et al. \cite{djelouah2019neural} introduce an autoencoder framework with a predictive coding scheme, using a motion estimation network to predict the next frame and a residual compression network to encode the prediction error. M-LVC \cite{lin2020m} leverages multiple reference frames to predict associated motion vectors and perform motion compensation for frame reconstruction in an end-to-end learned video compression scheme. C2F \cite{hu2022coarse} enhances motion compensation through a two-stage coarse-to-fine deep video compression framework and employs hyperprior-guided mode prediction for optimized block resolution and residual coding.

Unlike residual coding, conditional coding uses the reference frame as conditional information for transform and entropy coding, significantly advancing end-to-end video encoding. Ladune et al. \cite{ladune2020optical} show that encoding a video frame \(x_t\) with its motion-compensated reference frame \(x_c\) results in a lower entropy rate than encoding the residual signal \(x_t - x_c\) without conditions. Numerous studies \cite{liu2020conditional,li2021deep,li2022hybrid,li2023neural,li2024neural} explore using spatial-temporal context to implement conditional entropy models, thereby improving video compression performance. The DCVC \cite{li2021deep} series uses motion compensation to continuously obtain temporal context and improve encoding, decoding, and entropy coding. DCVC-HEM \cite{li2022hybrid} implements an entropy model that utilizes spatial-temporal context. DCVC-DC \cite{li2023neural} demonstrates rate-distortion performance exceeding the current best traditional video coding standard, VVC \cite{bross2021overview}. The latest work, DCVC-FM \cite{li2024neural}, optimizes the range of variable bitrates the model can achieve.

\subsection{INR-based Video Coding}

INR uses neural networks to represent data with two main paradigms for videos: data coordinates and autoregressive embeddings. The data coordinates methods \cite{chen2021nerv,li2022nerv,bai2023ps,gomes2023video,maiya2023nirvana} use neural networks to map pixel coordinates \((x,y,t)\) to corresponding pixel values \((R,G,B)\). However, the data coordinates do not contain any content information of the target data, making the expressive power of INR completely dependent on neural networks. The autoregressive embeddings methods \cite{chen2022cnerv,chen2023hnerv,lee2023ffnerv,tang2023scene,zhao2023dnerv,kwan2024hinerv} introduce embeddings updated during training to capture video content information, reducing network training difficulty and improving representation performance.

The emergence of INR provides a new approach for deep learning based video coding, which represents videos through a simple neural network and then compresses the neural network. NeRV \cite{chen2021nerv} first proposed an INR video compression pipeline consisting of video overfitting, model pruning, model quantization, and weight encoding. Subsequent methods such as E-NeRV \cite{li2022nerv}, HNeRV \cite{chen2023hnerv}, FFNeRV \cite{lee2023ffnerv}, etc. further improved the network structure based on NeRV \cite{chen2021nerv}, by introducing autoregressive embeddings, and adding optical flow references to enhance the representation performance of video INR, thereby improving the final video compression performance. Gomes et al. \cite{gomes2023video} improved the compression pipeline of NeRV \cite{chen2021nerv} by introducing network parameter information entropy loss during training to achieve joint rate-distortion optimization. Tang et al. \cite{tang2023scene} improved the performance of INR video representation by introducing optical flow supervision, frequency domain supervision, and contrastive loss. HiNeRV \cite{kwan2024hinerv} has divided autoregressive embeddings into different scales and improved the expressive power of the INR network under finite parameter conditions through parameter free upsampling. Zhang et al. \cite{zhang2024boosting} utilized a conditional decoder with a time aware affine transformation module and sine NeRV sample blocks to enhance the compression performance of INR videos.

\section{Proposed Method}
\subsection{Preliminaries}
We firstly introduce the structure and encoding pipeline of INR video compression based on the recent work, HiNeRV \cite{kwan2024hinerv}. Then, we analyze and design effective parameter reuse mechanism based on it.

\subsubsection{Network Structure}
\begin{figure*}[htbp]
    \centering
    \includegraphics[width=\textwidth]{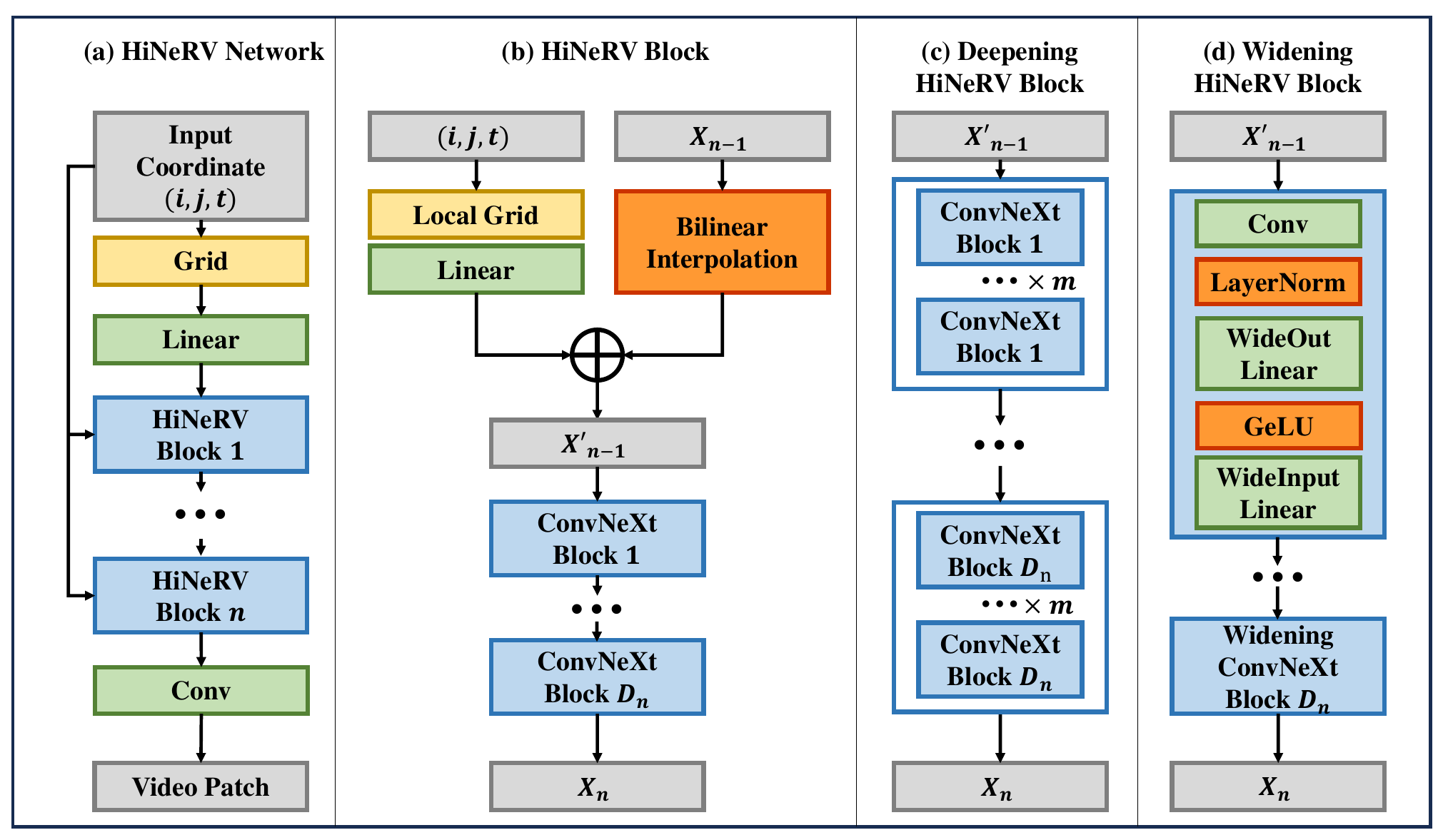}
    \caption{(a) shows the network structure of the HiNeRV network. The input is the target video patch coordinate (i,j,t). According to the coordinate, the input grid which is an autoregressive embedding is computed. After a linear layer, the input grid will be processed by n HiNeRV blocks. Finally, after a convlutional layer, the output video patch is generated. (b) gives the details of HiNeRV block. The block computes the local grid with the input coordinate (i,j,t) and maps it to the target channels by a linear layer. Then the local grid will add the upsampled input feature Xn-1 and be processed by Dn ConvNeXt blocks to get the output feature Xn. (c) shows the deepening HiNeRV block by parameter reuse. In the HiNeRV block, each ConvNeXt block will be stacked m times to reuse their parameters and enhance the network expressive ability. (d) shows the widening HiNeRV block with widening ConvNeXt block. In the widening ConvNeXt block, the first linear layer concats its weight to expand the output channel. The second linear layer concats its weight to expand the input channel.}
    \label{fig:framework}
\end{figure*}

As shown in Fig.\ref{fig:framework}(a), HiNeRV begins by mapping the input patch coordinates \((i,j,t)\) to a basic input feature \(X_0\):
\begin{equation}
    X_0 = F_{linear}(\gamma_{base}(i,j,t)),
\end{equation}
where \(\gamma_{base}\) represents the basic grid which is an autoregressive embedding in HiNeRV, and \(F_{linear}\) is a linear layer that adjusts the grid to the required number of channels. This initial feature \(X_0\) then enters $n$ HiNeRV blocks, where it is progressively processed and upsampled into the output video patch.

As shown in Fig.\ref{fig:framework}(b), each HiNeRV block performs a series of transformations, starting with bilinear interpolation to upsample the input feature \(X_{n-1}\) by a scale factor \(S_n\). It then calculates a hierarchical grid \(\gamma_{n}(i,j,t)\) based on the coordinate \((i,j,t)\) and maps it to the necessary number of channels through a linear layer. The upsampled feature \(X_{n-1}\) is then added with this processed grid. This mixed feature is fed into $D_n$ ConvNeXt blocks. A ConvNeXt block includes a depthwise separable convolution layer that extracts spatial information, followed by layer normalization. The features are then amplified by a linear layer, activated by a GeLU function, and reduced in channels by another linear layer to produce the final output.

Finally, the output of the last HiNeRV block is processed by a head layer that maps it to \((R,G,B)\) video frames using a convlutional layer. 

\begin{figure*}[htbp]
    \centering
    \includegraphics[width=\textwidth]{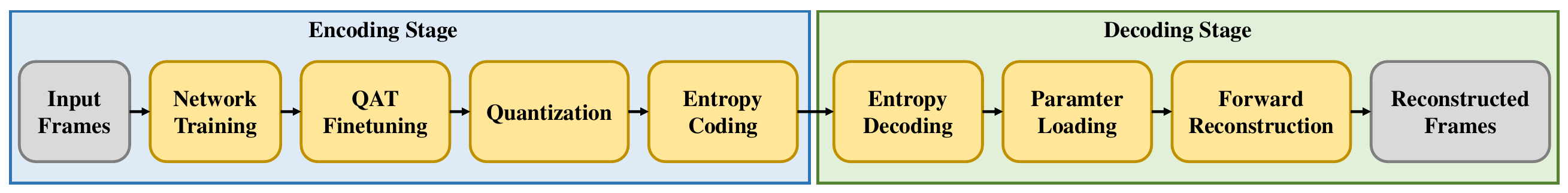}
    \caption{The encoding and decoding pipeline of our method.}
    \label{fig:pipeline}
\end{figure*}

\begin{table*}[htbp]
\centering
\begin{tabular}{cccccc}
\hline
& Basketball  & BQTerrace  & Cactus  & Kimono & ParkScene  \\ \hline
HiNeRV [bpp/PSNR]       & \begin{tabular}[c]{@{}c@{}} 0.0196 \\ 31.3108 \end{tabular} & \begin{tabular}[c]{@{}c@{}} 0.0170 \\ 31.6178 \end{tabular} & \begin{tabular}[c]{@{}c@{}}0.0196 \\ 31.7870\end{tabular} & \begin{tabular}[c]{@{}c@{}}0.0370 \\ 34.5848\end{tabular} & \begin{tabular}[c]{@{}c@{}}0.0371 \\ 31.6903\end{tabular} \\ \hline
Deepening HiNeRV [bpp/PSNR] & \begin{tabular}[c]{@{}c@{}} 0.0196 \\ 31.6158 \end{tabular} & \begin{tabular}[c]{@{}c@{}} 0.0171 \\ 31.7672\end{tabular} & \begin{tabular}[c]{@{}c@{}}0.0196 \\ 32.0939\end{tabular} & \begin{tabular}[c]{@{}c@{}}0.0371 \\ 34.8971\end{tabular} & \begin{tabular}[c]{@{}c@{}}0.0372 \\ 32.0078\end{tabular} \\ \hline
Widening HiNeRV [bpp/PSNR]  & \begin{tabular}[c]{@{}c@{}}0.0196 \\ 31.3125\end{tabular} & \begin{tabular}[c]{@{}c@{}}0.0171 \\ 31.6143\end{tabular} & \begin{tabular}[c]{@{}c@{}}0.0197 \\ 31.8133\end{tabular} & \begin{tabular}[c]{@{}c@{}}0.0371 \\ 34.5866\end{tabular} & \begin{tabular}[c]{@{}c@{}}0.0372 \\ 31.6721\end{tabular} \\ \hline
\end{tabular}
\caption{ The parameter reusing tests on HEVC Class B are shown in the table. Deepening HiNeRV means stacking the ConvNeXt block for once. Widening HiNeRV means the linear layer in HiNeRV block concat its weight for once.}
\label{reuse test}
\end{table*}

\subsubsection{Encoding and Decoding}
In this paper, we train the HiNeRV network using mean-square-error (MSE) as the loss function to represent the target video. We train the network and makes it represent the target video effectively. After achieving initial trained network, we enhance the network's performance through quantization-aware-training (QAT). QAT introduces quantization noise as a substitute for direct quantization to prevent the zero-gradient problem. This approach allows us to quantize the network parameters to 6 bits without losing significant information. During QAT, we finetune the network to maintain high performance despite the reduced precision. This step is crucial for achieving efficient compression while preserving video quality. We then quantize the network parameters with 6 bit and apply arithmetic coding to the quantized network parameters, transforming them into a compact bitstream suitable for transmission and storage.

The decoding process begins with arithmetic decoding, which retrieves the quantized network parameters. These parameters are then reloaded into the HiNeRV network. By performing forward propagation, the network reconstructs the video frames corresponding to each input coordinate. 


\subsection{INR Video Compression Potential}

The primary objective of video compression is to minimize the data size while maintaining high video quality or to enhance the quality of the reconstructed video within a fixed data size. For INR video compression, this challenge becomes an optimization problem that balances the number of parameters and video quality. Despite the already minimal amount of parameters in current INR models and the use of network compression techniques aimed at further reducing data size, we suggest a new angle: \textbf{“Has the information storage potential of INR network parameters been fully utilized?”}


To explore whether INR network parameters fully exploit their information storage potential, we examine the development trajectory of INR video compression. Initially, the NeRV \cite{chen2021nerv} model uses simple CNNs to map coordinates to video frames, then fits the video and compresses the network for video compression. Subsequent efforts make minimal changes to video fitting and network compression, instead focusing on improving network architecture. Innovations like autoregressive embedding \cite{chen2023hnerv}, complex ConvNext blocks \cite{kwan2024hinerv}, and auxiliary encoders \cite{zhang2024boosting} significantly enhance the reconstruction quality of INR models with the same number of parameters. These advancements demonstrate that architectural improvements effectively increase the information storage capacity of INR parameters.


Given this background, it is reasonable to assume that the information storage potential of INR network parameters has not been fully realized. To explore this further, we propose a method that focuses on enhancing the utilization of existing network parameters through parameter reuse, rather than altering or adding new INR structures. Traditionally, a network's expressive power increases with deeper and wider architectures, which often results in a larger amount of parameters. However, from a compression perspective, we hypothesize that it is possible to increase the network's depth and width without increasing the amount of parameters to be encoded. By reusing parameters, we aim to enhance the network's expressive capability while maintaining parameter amount. This approach challenges the conventional trade-off between complexity and parameter amount, suggesting that parameter reuse can achieve the desired expressive power without additional parameters to be encoded. Our method seeks to maximize the potential of current architectures.

We implement parameter reuse in the ConvNeXt blocks of the HiNeRV network. This approach, depicted in Fig.\ref{fig:framework}(c), augments the network's depth while keeping the amount of encoded parameters unchanged. Firstly, we only reuse the ConvNeXt block for once, and evaluate the compression performance on the HEVC Class B dataset. The findings, presented in Table \ref{reuse test}, reveal that the upgraded HiNeRV network with parameter reuse exceeds the performance of the original network obviously. This demonstrates that the parameter reuse strategy effectively unleashes the parameter potential of neural representation for efficient video compression.

\subsection{Parameter Reuse}
Our proposed parameter reuse mechanism aims to enhance the expressive capability of the HiNeRV network by reusing parameters within its ConvNeXt blocks. This approach allows for deepening and widening the network without increasing the overall parameter amount. Below is a detailed description of our parameter reuse mechanism.

\subsubsection{Deepening the Network}

\textbf{Stacking ConvNeXt Blocks:} We implement a scheme where ConvNeXt blocks are repeatedly stacked. Specifically, we repeat the layer structure of a ConvNeXt block $m$ times as shown in Fig.\ref{fig:framework}(c). Each stacked ConvNeXt block uses the same set of parameters, which deepens the network without increasing the parameter amount.

\textbf{Implementation Steps:} During the forward pass, for each input $X$, the output $Y$ is obtained after passing through $m$ identical ConvNeXt blocks. The process is represented as follows:
\begin{equation}
     Y = ConvNeXt^m(X).
\end{equation}
This method increases the network's depth while keeping the amount of encoded parameters unchanged since each layer reuses the same parameters. We have also conducted more exploration and experiments on the location, number of layers, and granularity of parameter reuse deepening. The specific results are shown in the experimental section.

\subsubsection{Widening the Network}
\textbf{Concatenation of Linear Layer Weights:} Because when widening a neural network, we need to consider aligning the number of input and output channels. Therefore, we consider expanding the output channels of the first linear layer and the input channels of the second linear layer in ConvNeXt block, ensuring that the input and output channels of each ConvNeXt block are aligned. To widen the linear layers, we concatenate the weights of two linear layers in the ConvNeXt blocks. This means that the weight matrices of these layers are repeated and concatenated to form a wider matrix without adding new parameters.

\textbf{Implementation Steps:} For an input \( X \), we create a new weight matrix by concatenating the original weight matrix. These concatenated weights process the input, thereby widening the network.
\begin{align}
    W_{1\text{new}} &= \text{concat}(W_1, W_1, \text{axis}=0), \\
    W_{2\text{new}} &= \text{concat}(W_2, W_2, \text{axis}=1),
\end{align}
where $W_1$ and $W_2$ represent the original weight matrixes, and $W_{1new}$ and $W_{2new}$ are the new concatenated weight matrixes. The output \( Y \) is obtained by applying the two wider linear operations with $W_{1new}$ and $W_{2new}$.
\begin{equation}
 Y = Conv(X)W_{1new}^{T}W_{2new}^{T}.
\end{equation}
This concatenation mechanism increases the network's width while maintaining the same parameter count.

We also conduct tests for the network widening method on HEVC Class B. And the video compression results are shown in Table.\ref{reuse test}. The test results show that widening the network does not improve the video compression performance. The scheme of widening the network through parameter reuse cannot stimulate the potential of the network to preserve information, so we do not widen the network again in our subsequent experiments.

Although widening the network cannot improve the video compression performance, deepening the network is obviously benefit to compression performance. Therefore, parameter reuse enables a smaller set of parameters to perform video representation task, improving the parameter utilization efficiency. The reuse mechanism allows the network to deepen its structure, enhancing its ability to model complex patterns in video data. This method improves reconstruction quality while keeping the encoded parameter amount stable.

\begin{figure*}[htbp]
    \centering
\begin{tabular}{@{} c @{}}

  \begin{tabular}{@{} c @{} c @{}}

    \includegraphics[width=0.5\textwidth]{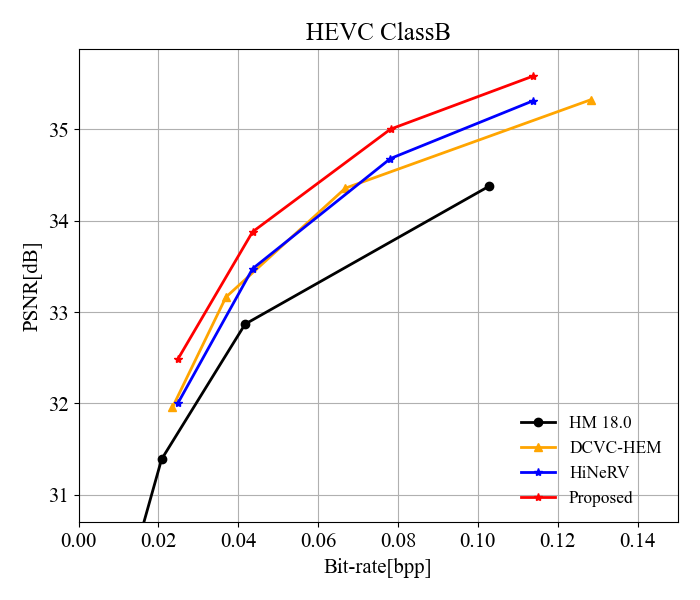} &
    \includegraphics[width=0.5\textwidth]{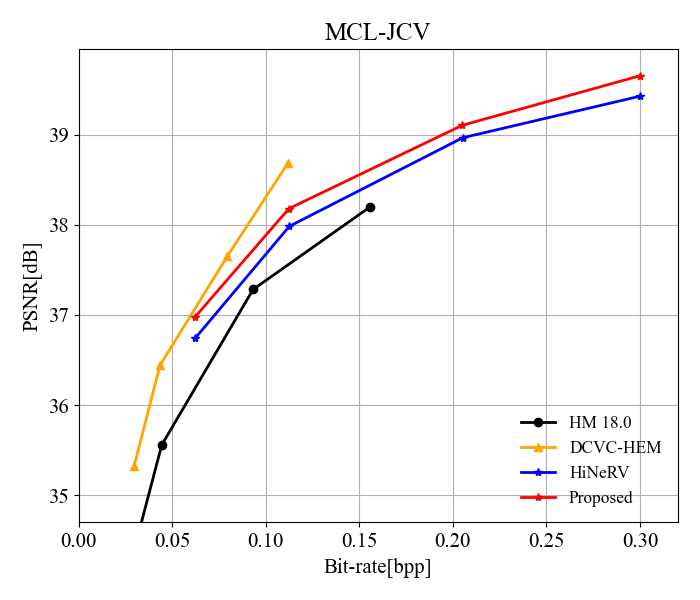}

  \end{tabular}
\end{tabular}

\caption{The rate-distortion performance of our method and other video compression methods on HEVC Class B and MCL-JCV datasets.}
    \label{fig:rd}
\end{figure*}

\begin{table*}[htbp]
\centering
\begin{tabular}{cccc}
\hline
              & HEVC                                                & DCVC-HEM                                                 & HiNeRV                                                                                \\ \hline
HEVC ClassB        & \begin{tabular}[c]{@{}c@{}} -37.16\%  \end{tabular} & \begin{tabular}[c]{@{}c@{}} -16.66\%  \end{tabular} & \begin{tabular}[c]{@{}c@{}}-16.34\% \end{tabular} \\ \hline
MCL-JCV & \begin{tabular}[c]{@{}c@{}}-26.20\% \end{tabular} & \begin{tabular}[c]{@{}c@{}}12.01\% \end{tabular} & \begin{tabular}[c]{@{}c@{}}-10.83\% \end{tabular} \\ \hline
\end{tabular}
\caption{The BD-rate benefits of our method against other methods on HEVC Class B and MCL-JCV datasets.}
\label{bd rate}
\end{table*}

\section{EXPERIMENT}\label{experiment}

\subsection{Experiment setting}

\subsubsection{Training setting}
﻿We use PyTorch to build and train our network. For optimization, we choose the Adam \cite{kingma2014adam} optimizer. We initialize the learning rate at 5e-4 and use a cosine annealing learning rate schedule \cite{loshchilov2016sgdr}. We set training epochs of 300, warmup epochs of 30, and QAT epochs of 30. We utilize a single V100 GPU with 32GB memory to facilitate the training process.

\subsubsection{Test datasets}
We conduct tests on two common video datasets, HEVC Class B and MCL-JCV \cite{wang2016mcl}. HEVC Class B contains five 1920 × 1080 video sequences, totally 2080 frames. MCL-JCV contains 30 1920 × 1080 video sequences, totaling 4500 frames.

\subsubsection{Method configurations}
In our evaluation, we compare the performance of our proposed method for INR video compression with several state-of-the-art methods that represent recent trends in the field. These methods include:
\begin{itemize}
    \item HM: We choose HM18.0 as the representative of traditional methods. We take the default Main Profile with low-delay-p as the encoding config.
    \item DCVC-HEM: DCVC series are the representatives of end-to-end video encoding and have the best rate-distortion performance. We adopt the same low-delay-p setting as HM18.0 for all video sequences.
    \item HiNeRV: HiNeRV is the state-of-the-art INR-based video compression method. We set the depth to \{3,3,3,1\} and channels to \{280,400,560,688\} for different rates.
    \item  Proposed: We take the same setting for HiNeRV. For the ConvNeXt block reusing times, we set it to 2.
\end{itemize}

\begin{figure*}[htbp]
    \centering
\begin{tabular}{@{} c @{}}

  \begin{tabular}{@{} c @{} c @{} c @{} c @{} c @{}}

    \includegraphics[width=0.2\textwidth]{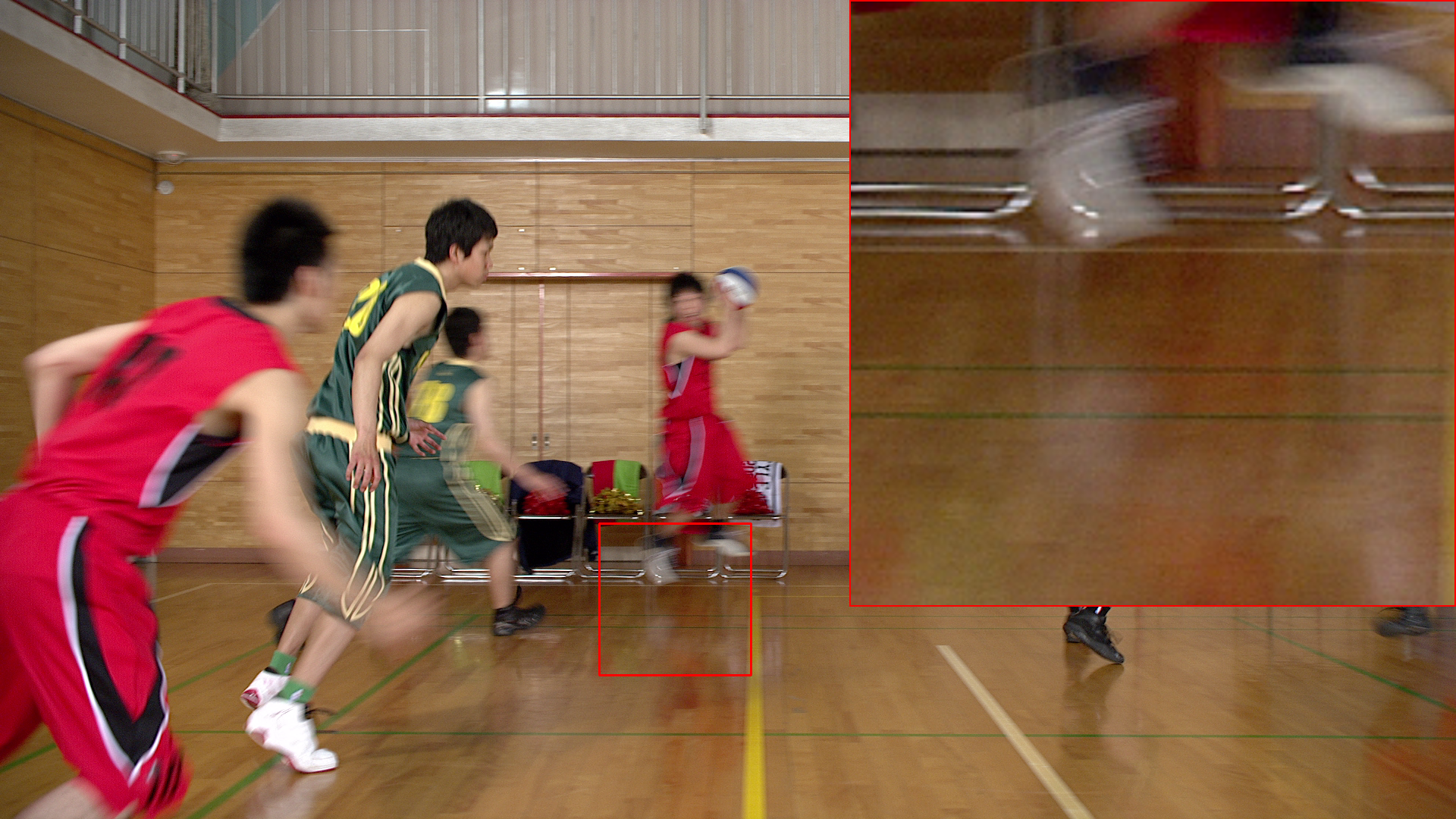} &
    \includegraphics[width=0.2\textwidth]{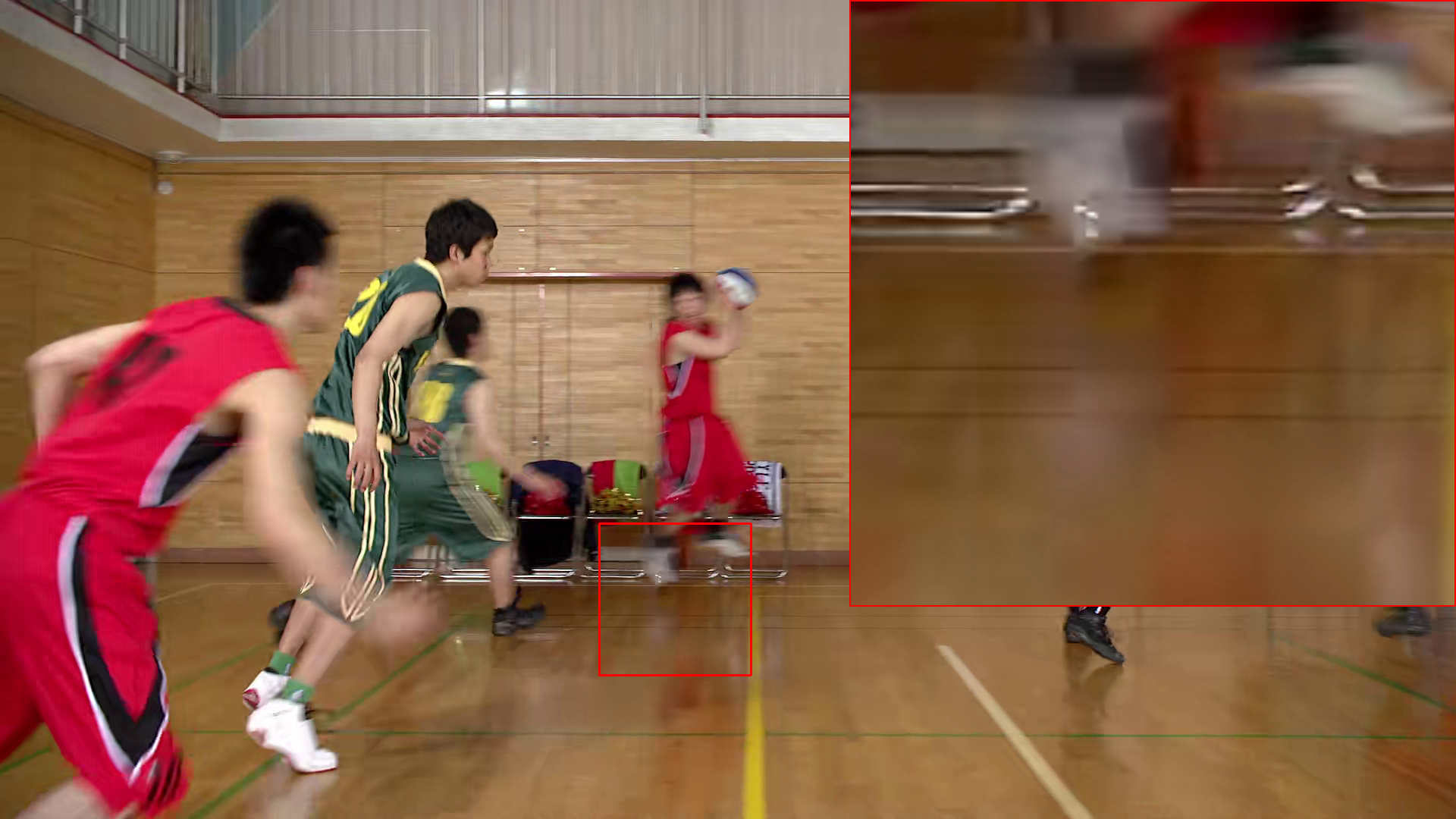} &
    \includegraphics[width=0.2\textwidth]{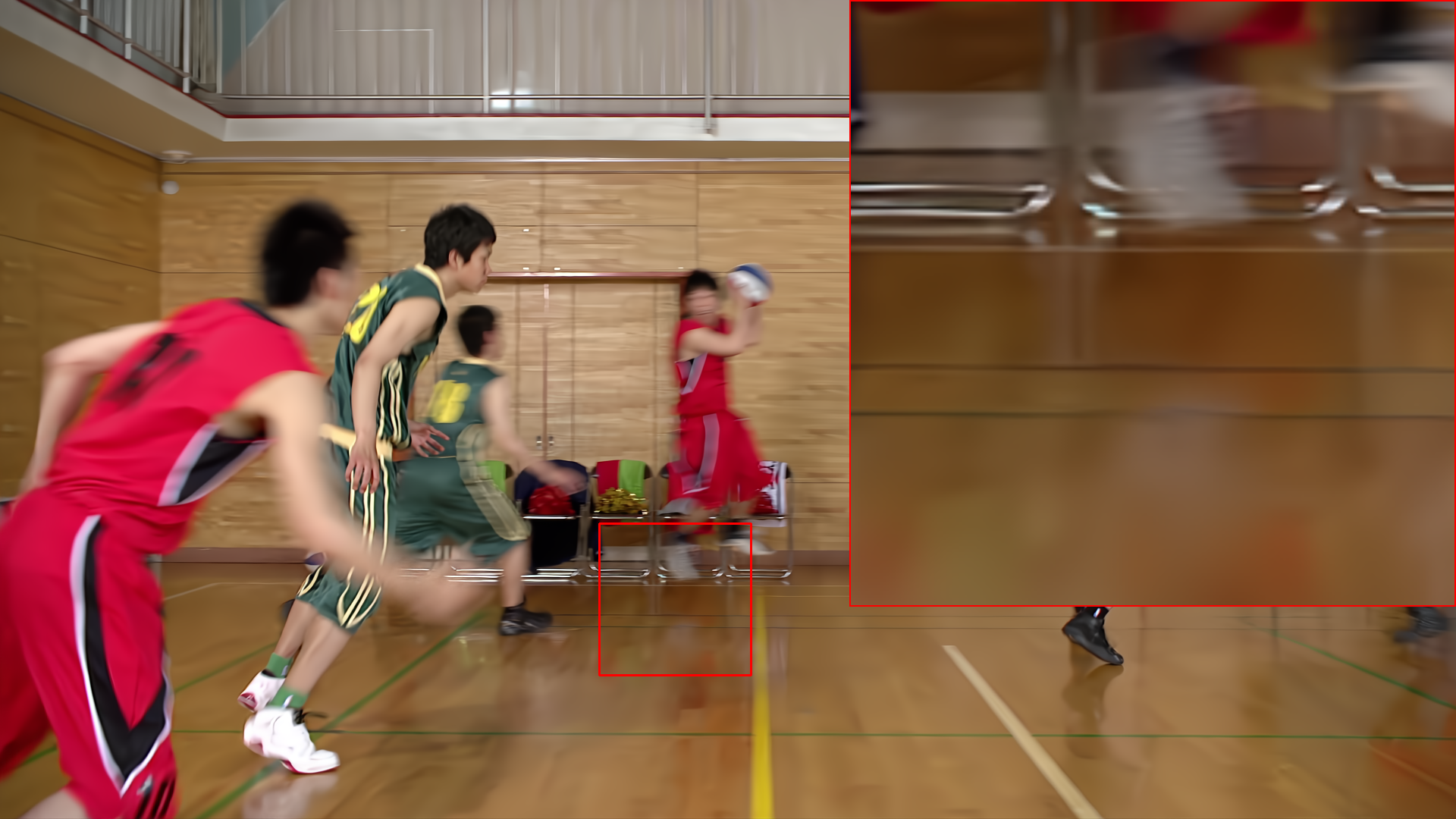} &
    \includegraphics[width=0.2\textwidth]{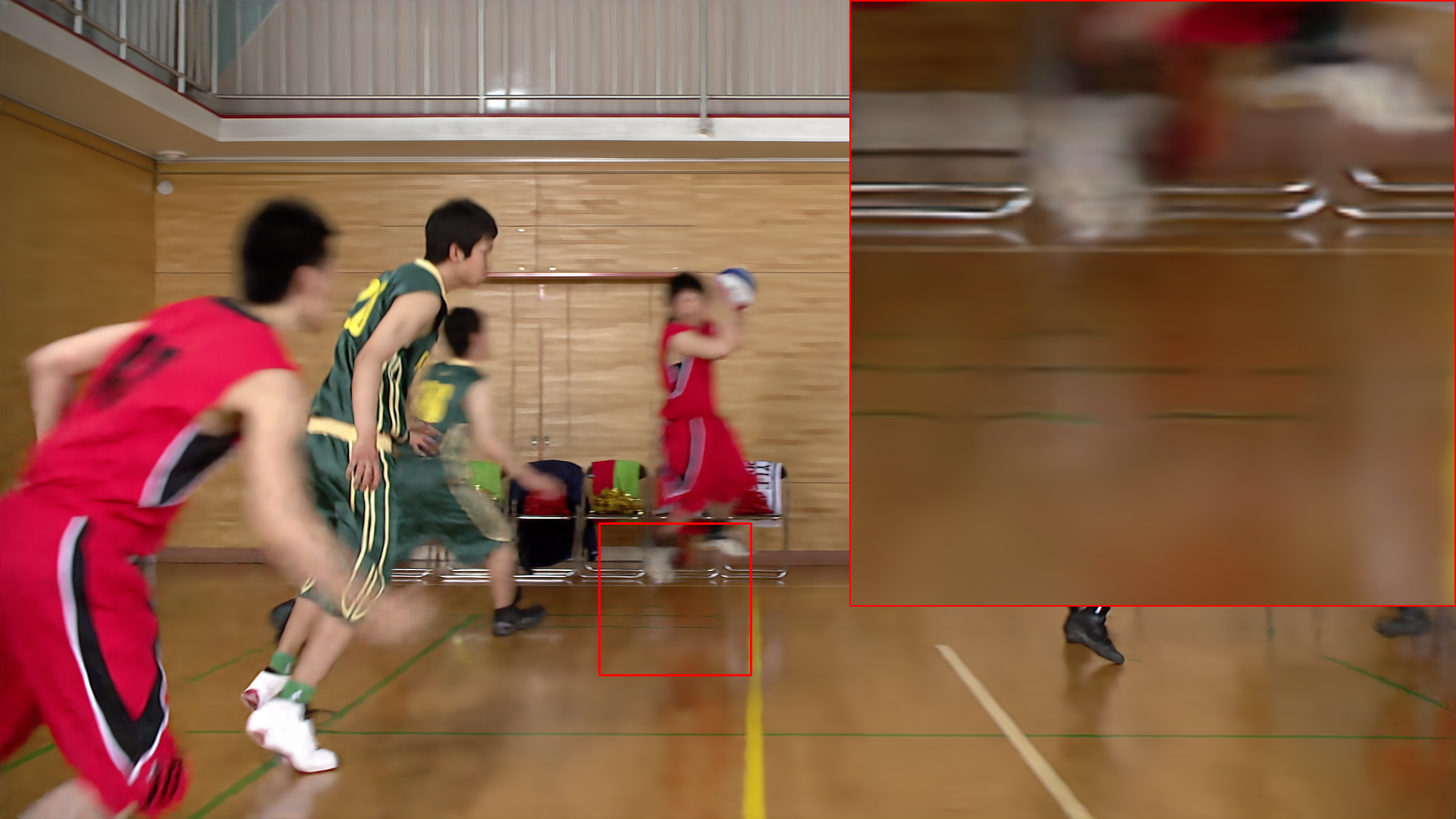} &
    \includegraphics[width=0.2\textwidth]{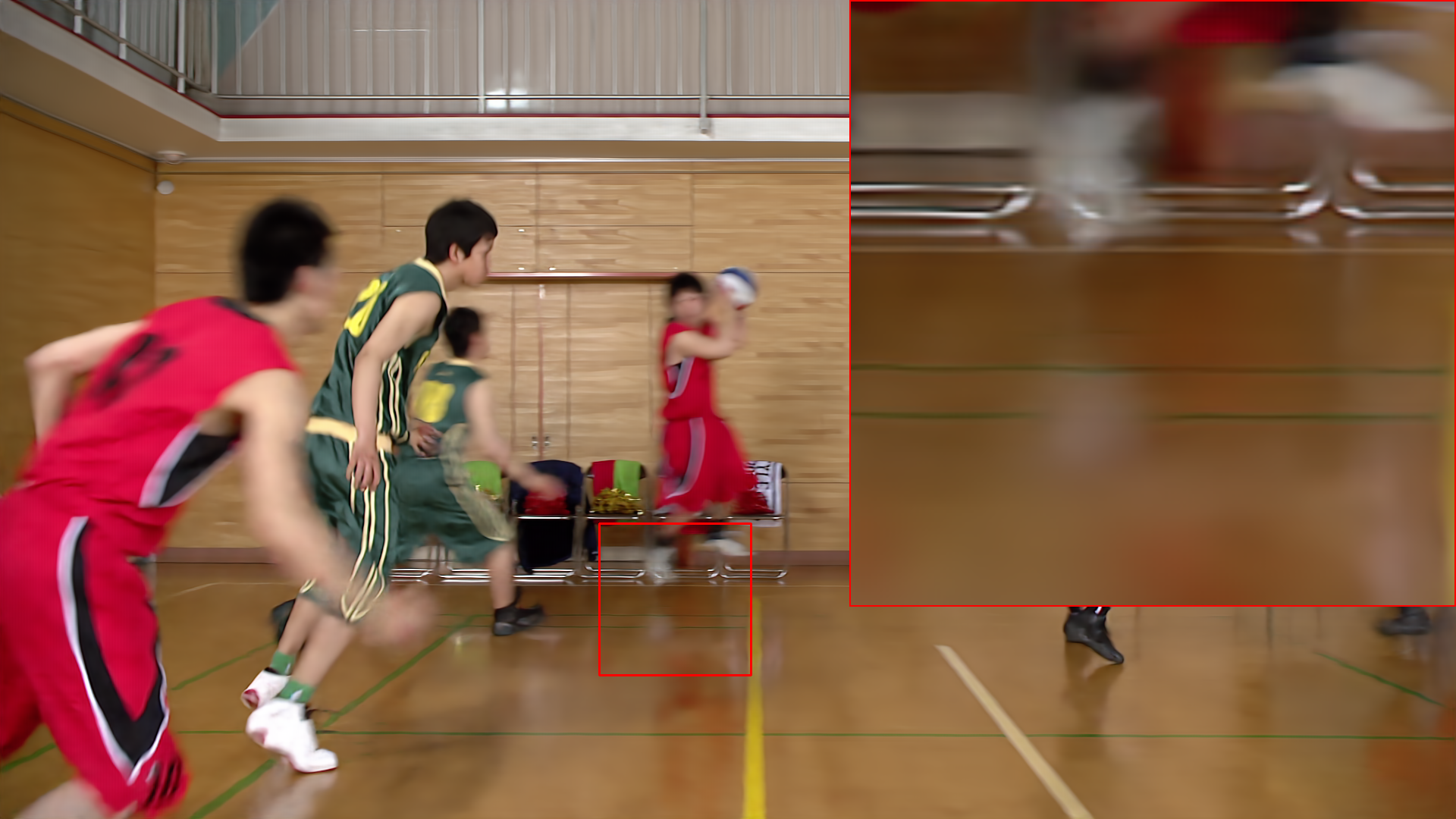} 
    \\

    GT & HM18.0 & DCVC-HEM & HiNeRV & Proposed \\
     & 31.99dB 0.0265bpp & 34.00dB 0.0384bpp & 32.60dB 0.032bpp & 33.06dB 0.032bpp \\

    \includegraphics[width=0.2\textwidth]{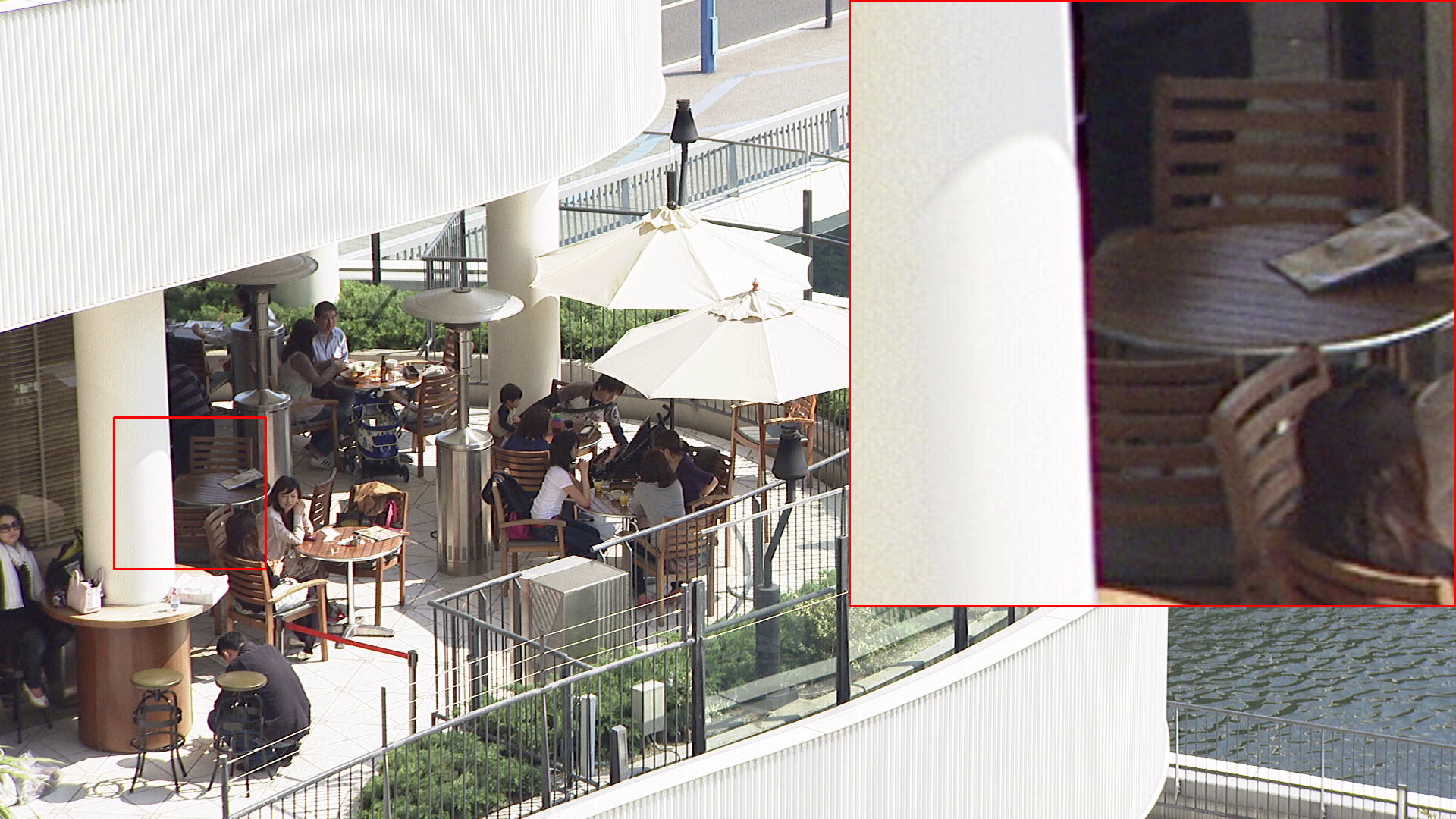} &
    \includegraphics[width=0.2\textwidth]{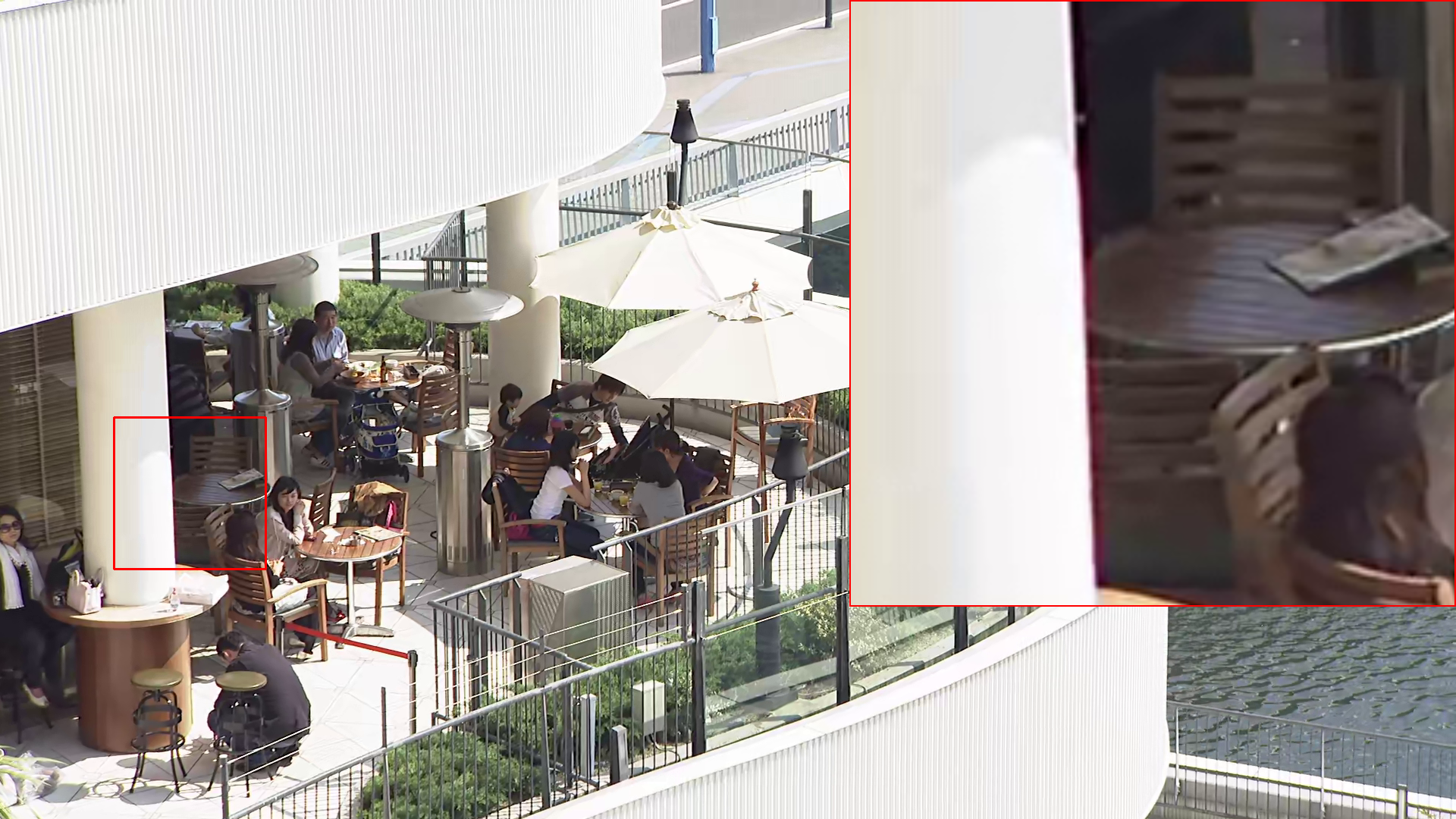} &
    \includegraphics[width=0.2\textwidth]{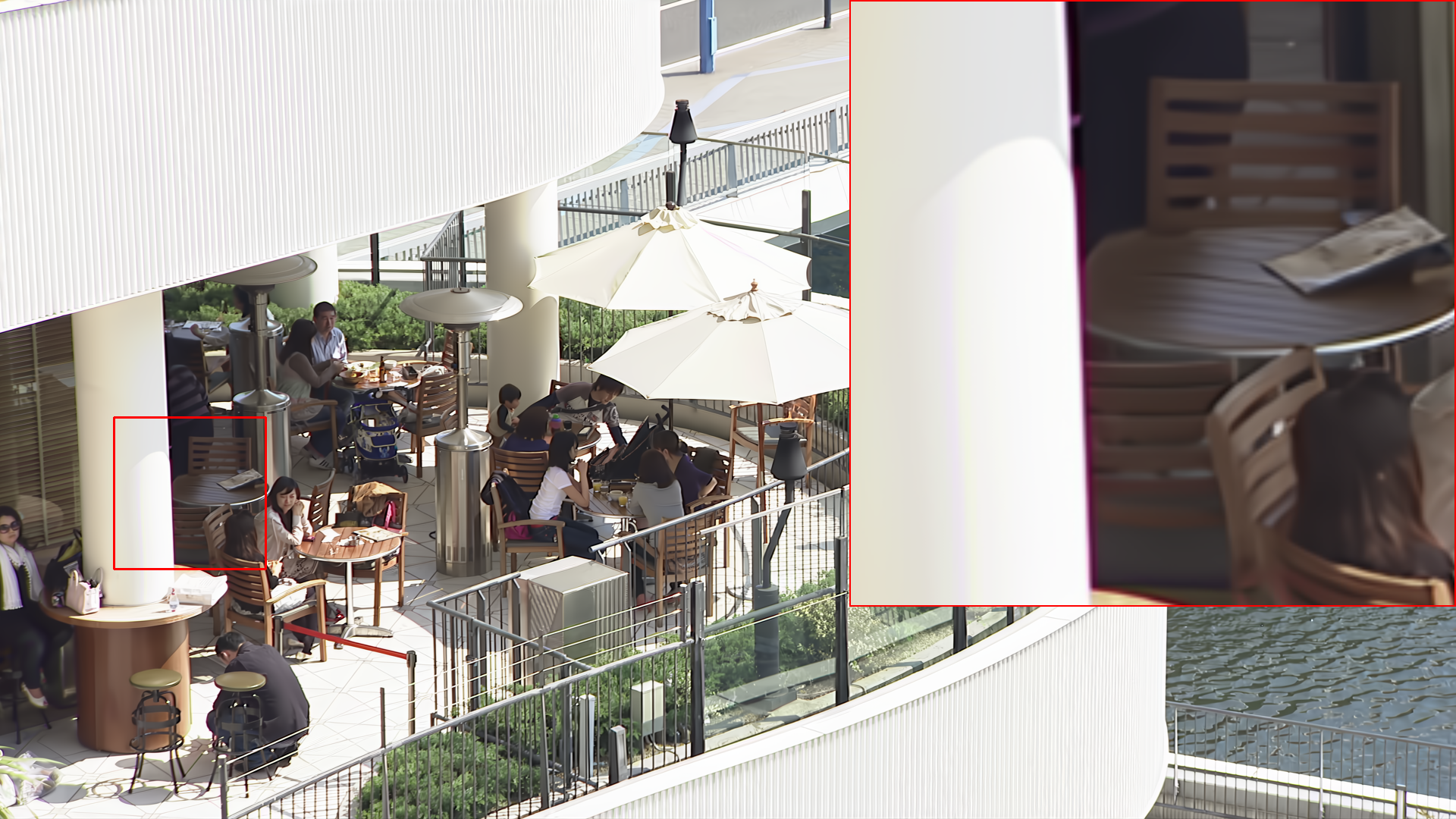} &
    \includegraphics[width=0.2\textwidth]{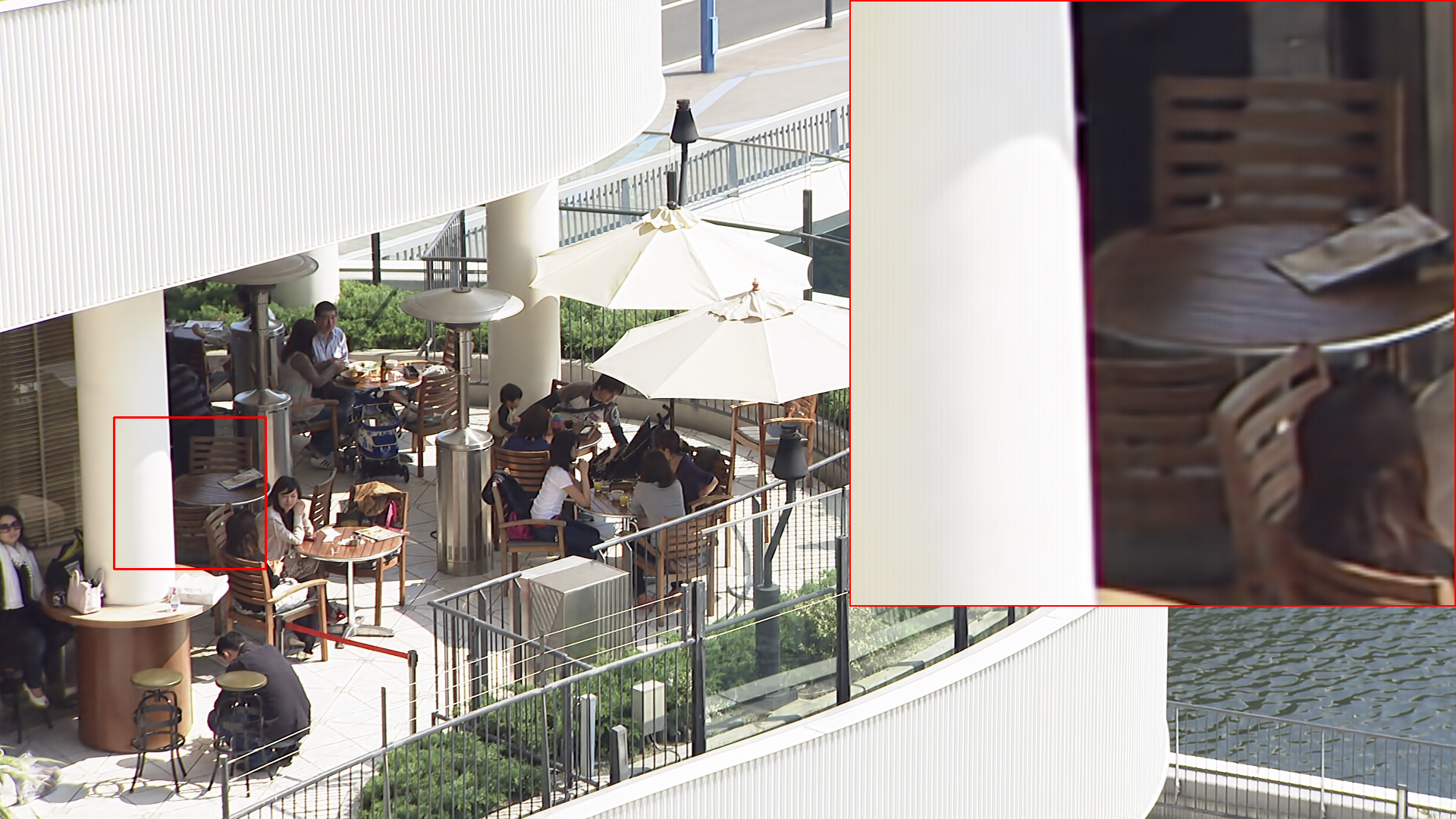} &
    \includegraphics[width=0.2\textwidth]{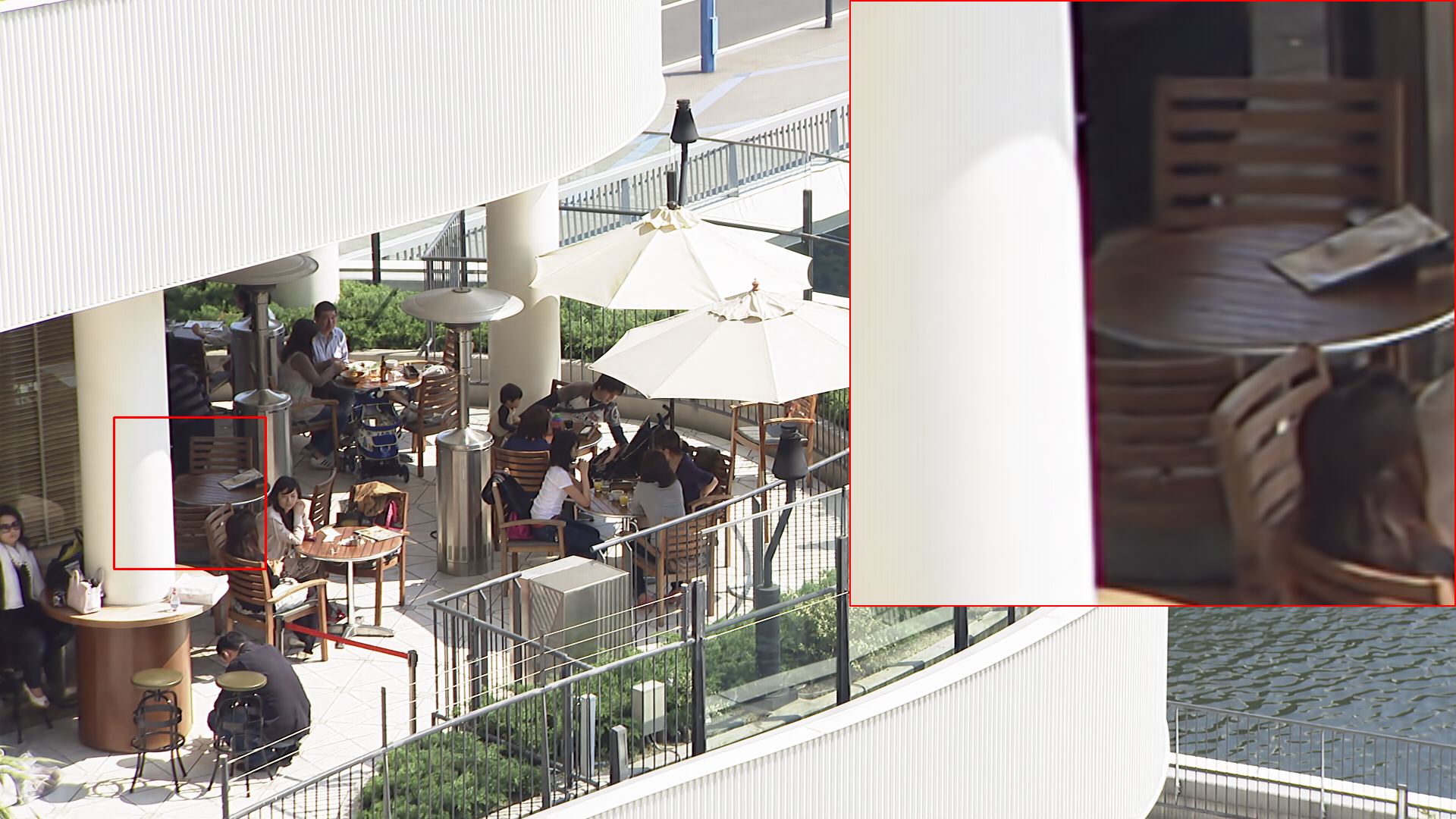} 
    \\

    GT & HM18.0 & DCVC-HEM & HiNeRV & Proposed \\
     & 30.43dB 0.0143bpp & 31.34dB 0.036bpp & 32.58dB 0.0273bpp & 32.82dB 0.0273bpp \\

    \includegraphics[width=0.2\textwidth]{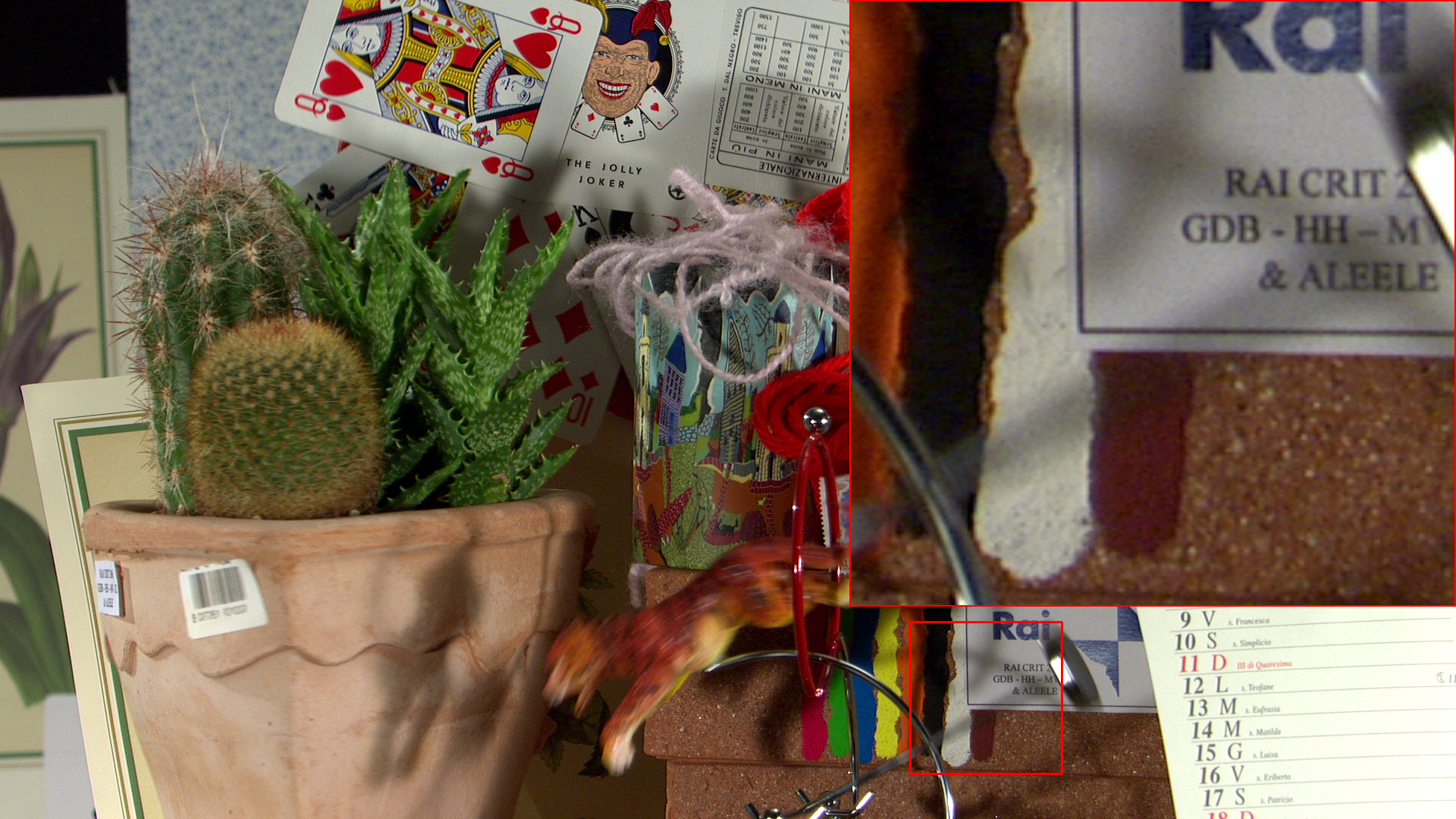} &
    \includegraphics[width=0.2\textwidth]{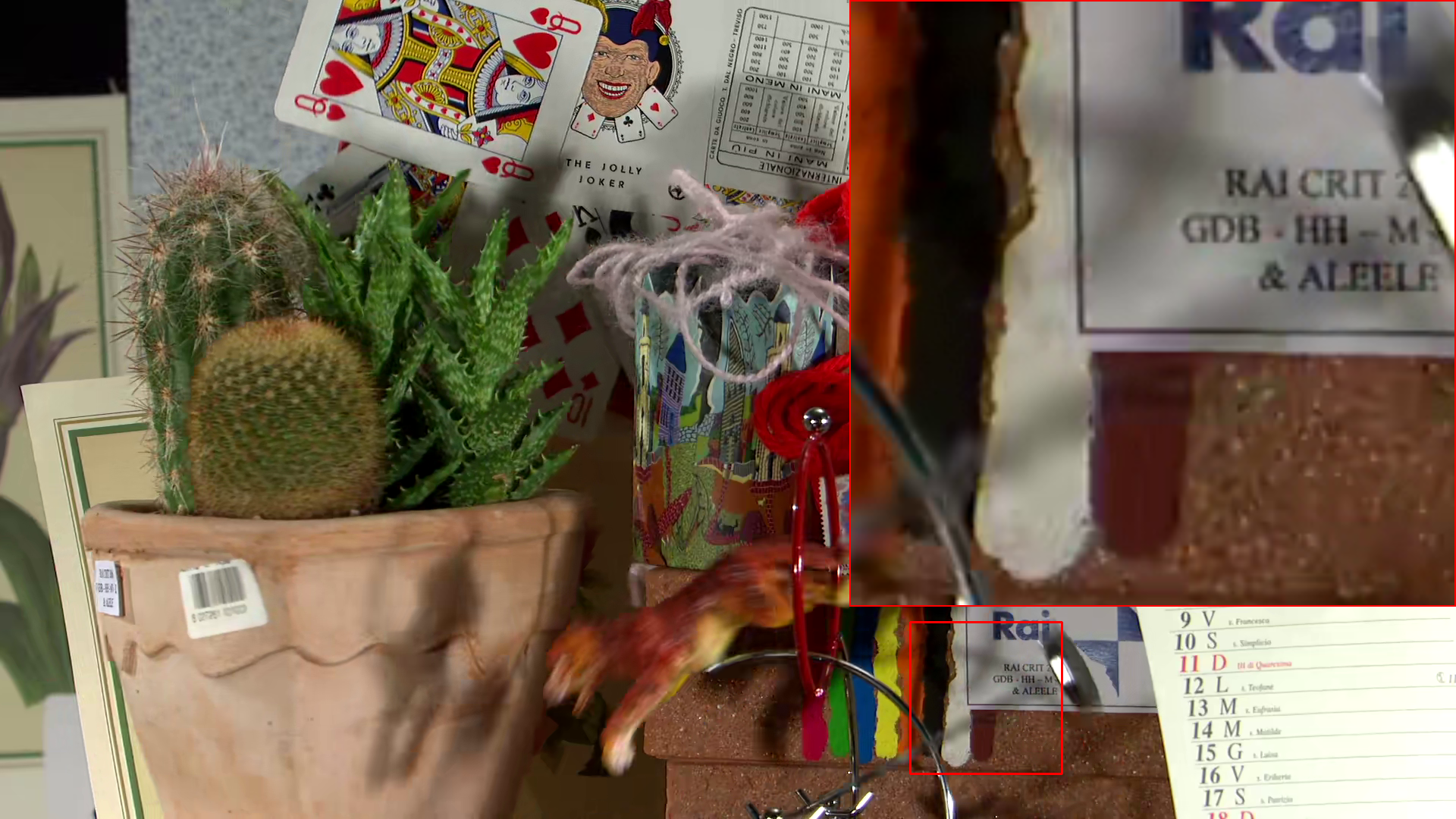} &
    \includegraphics[width=0.2\textwidth]{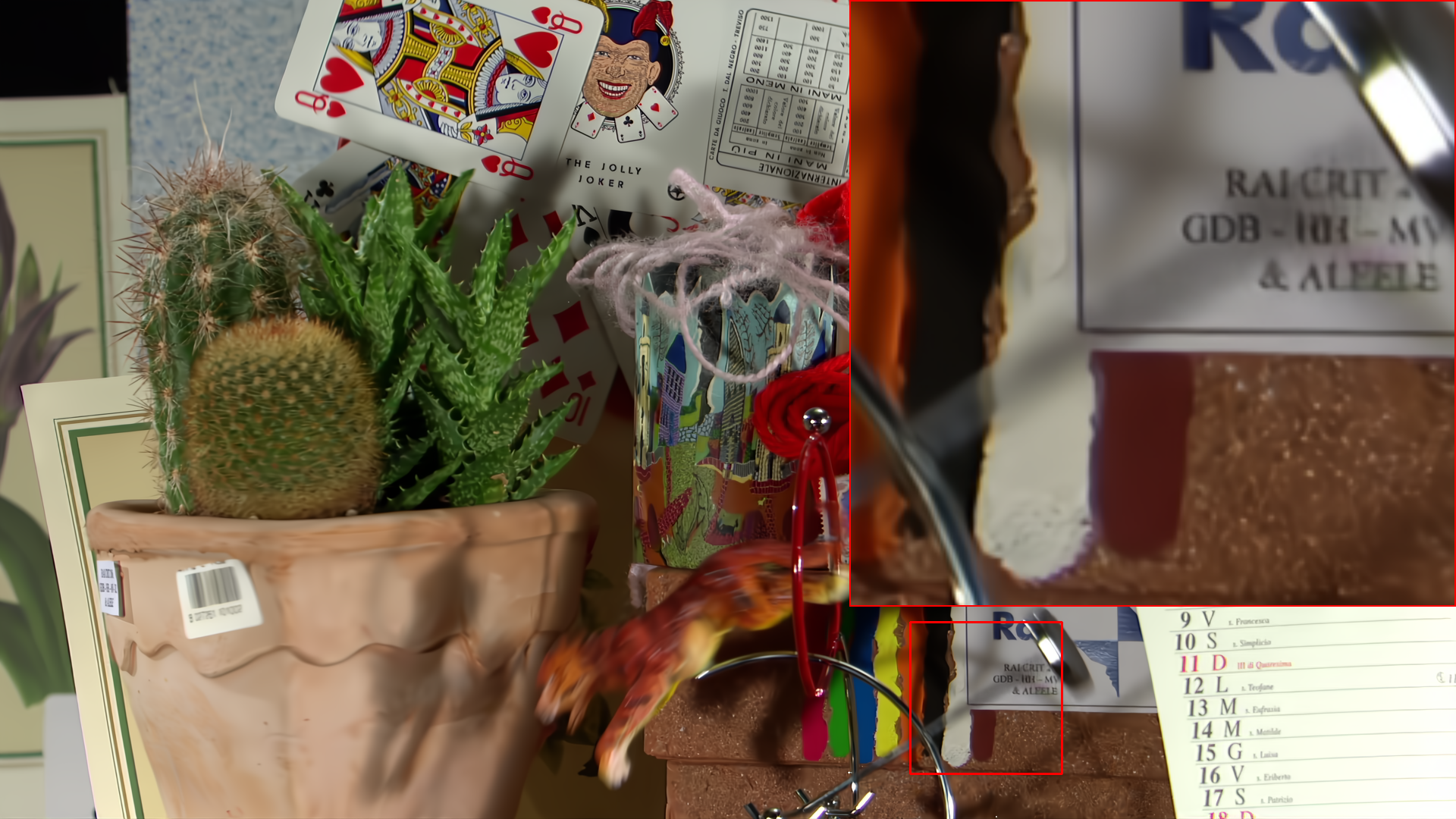} &
    \includegraphics[width=0.2\textwidth]{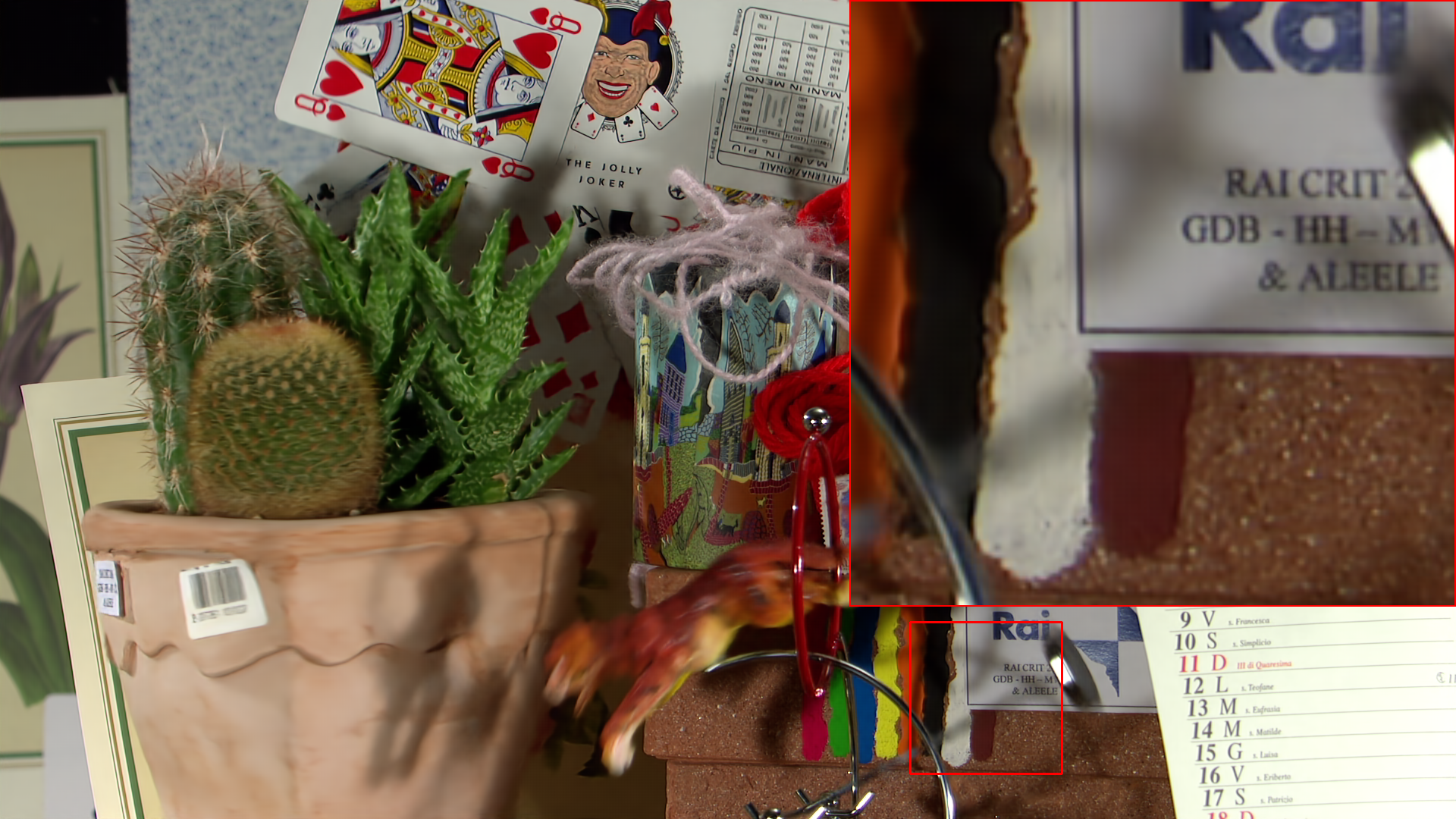} &
    \includegraphics[width=0.2\textwidth]{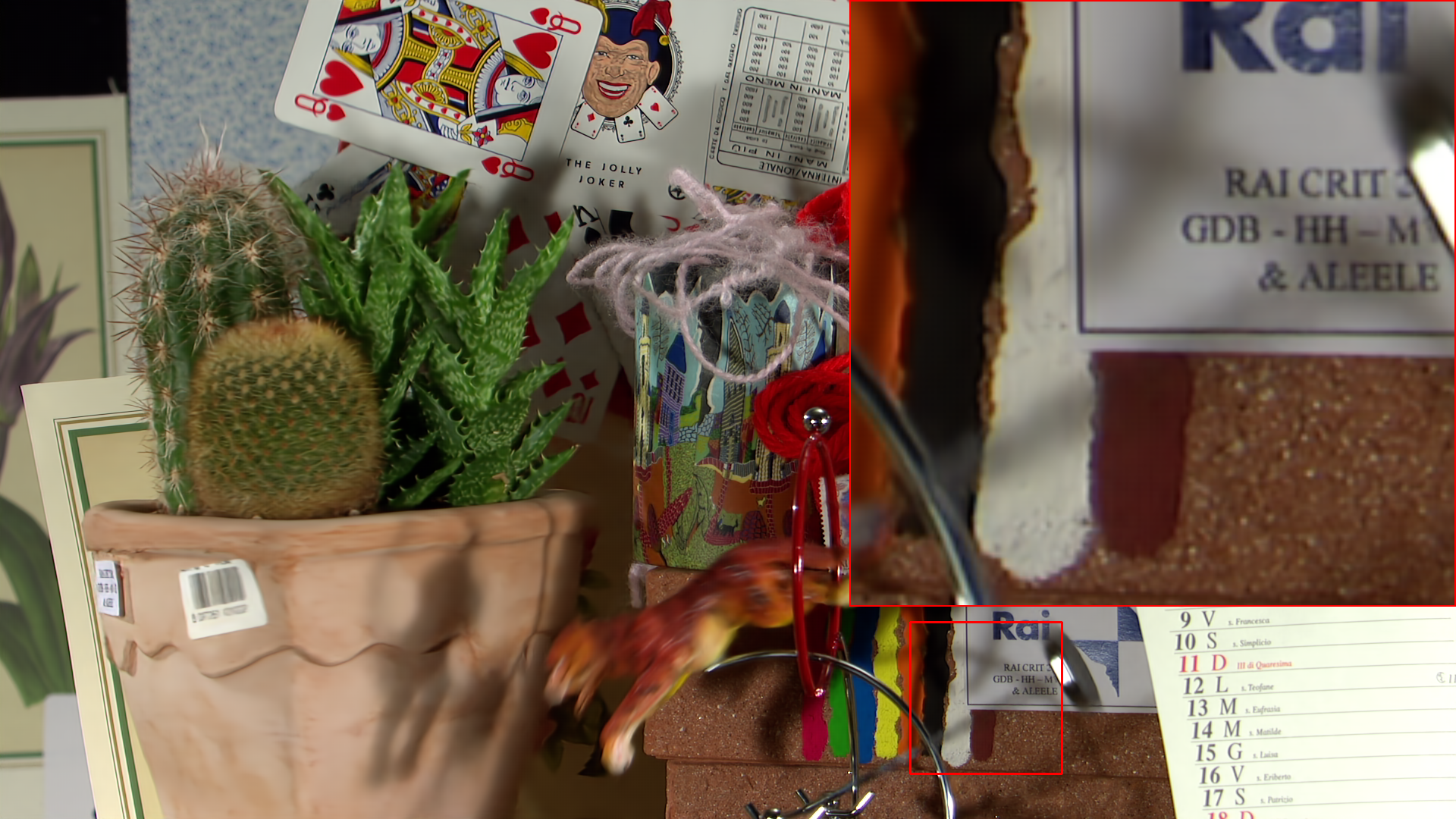} 
    \\

    GT & HM18.0 & DCVC-HEM & HiNeRV & Proposed \\
     & 31.06dB 0.021bpp & 32.31dB 0.036bpp & 33.04dB 0.032bpp & 33.40dB 0.032bpp \\

    \\

  \end{tabular}
\end{tabular}

\caption{The visual frames from HEVC Class B.}
    \label{fig:visual}
\end{figure*}

\subsection{Main Video Compression Results}
\subsubsection{Quantitative Results}
We use bpp (bits per pixel) as a measure of video compression rate and PSNR (Peak Signal to Noise Ratio) as a measure of reconstructed video quality. Fig.\ref{fig:rd} displays the rate-distortion performance curves for the HEVC Class B and MCL-JCV datasets. Table \ref{bd rate} presents the average BD-rate results for these datasets. On HEVC Class B, our method surpasses all other methods, including DCVC-HEM, a leading end-to-end video compression method. Notably, this marks the first time that an INR video compression method outperforms DCVC-HEM on HEVC Class B. On MCL-JCV, while our method does not exceed DCVC-HEM, it still performs better than HM18.0. Additionally, our method outperforms the INR video compression method, HiNeRV, on both datasets. The consistent improvement over HiNeRV across both datasets indicates that our method effectively activates HiNeRV's potential for storing information. By reusing parameters to deepen the network, our method enhances the expressive capability of INR without increasing network parameters, thereby optimizing rate-distortion performance. This strategy offers a new direction for improving INR video compression, alongside existing methods such as quantization-aware training with entropy constraints and more effective INR design.

\subsubsection{Visual Results}
We present three subjective images from the HEVC Class B dataset, each at a comparable bitrate. Our method shows clear advantages across different sequences. In \textit{BasketballDrive}, it distinctly outlines the stadium sideline with much greater clarity. In \textit{BQTerrace}, it captures the texture on the back of the chair with enhanced precision, making details more visible. In \textit{Cactus}, our method renders text that is sharper and more legible. These examples highlight our method's superior ability to maintain visual fidelity at similar bitrates.

\subsection{Decoding Complexity}
In terms of computational complexity, with the same depth of \{3,3,3,1\} and a channel size of 280, HiNeRV requires 181.89 GMACs for a sequence with 240 frames at 1920×1080 resolution. When using one ConvNeXt block of parameter reuse, the complexity increases to 291.36 GMACs, and with two ConvNeXt blocks of parameter reuse, it rises to 400.83 GMACs. Our method employs two ConvNeXt blocks of parameter reuse, doubling HiNeRV's MACs. As discussed in the method section, parameter reuse balances computational complexity and parameter quantity to achieve enhanced performance. Our ablation studies demonstrate that parameter reuse at different locations improves rate-distortion performance while maintaining acceptable computational complexity.

\begin{figure}[htbp]
    \centering
    \includegraphics[width=\columnwidth]{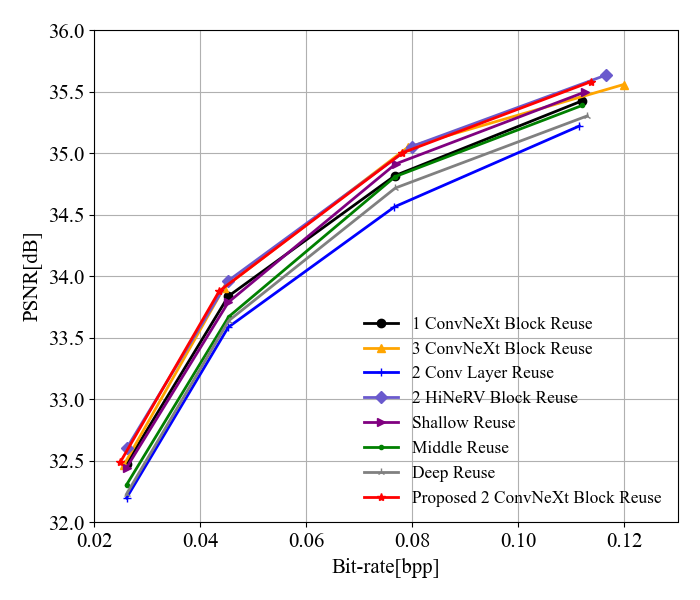}
    \caption{The ablation study on HEVC Class B of our method.}
    \label{fig:ablation}
\end{figure}

\subsection{Ablation Studies}
\subsection{Reuse location}
In our experimental setup, we define the HiNeRV blocks with a depth configuration of \{3,3,3,1\}, totaling four blocks. The last HiNeRV block, due to the unequal number of input and output channels, does not allow for parameter reuse. Therefore, we focus on the first three blocks, categorizing them as shallow, medium, and deep layers based on their proximity to the input coordinates, and apply two ConvNeXt blocks of parameter reuse on each. The results, shown in Fig.\ref{fig:ablation}, indicate that the shallow layer achieves the best performance with 236.3 GMACs on a sequence of 240 frames at 1920×1080 resolution. The mid-level reuse shows 263.83 GMACs, while the deep reuse offers minimal performance improvement, registering 264.48 GMACs. Considering the balance between computational complexity and performance, parameter reuse proves most effective in the shallow and middle layers.

\subsubsection{Reuse times}
We analyze the impact of twice parameter reuse in the main experiments. As Fig.\ref{fig:ablation} shows, we compare the rate-distortion effects of one ConvNeXt block reuse and three ConvNeXt block parameter reuse. Two ConvNeXt block parameter reuse outperforms once reuse and shows similar results to three times reuse, suggesting that two ConvNeXt block parameter reuse represents the upper limit for optimal performance.

\subsubsection{Reuse granularity}
We explore parameter reuse at different granularities. Fine-grained reuse targets two depthwise separable convolution layer reuse within ConvNeXt blocks, while coarse-grained reuse involves two HiNeRV block reuse. As shown in Fig.\ref{fig:ablation}, two convlutional layer reuse proves less effective compared to two ConvNeXt blocks of parameter reuse. However, two HiNeRV block reuse yields similar results to two ConvNeXt blocks of parameter reuse.

\section{Conclusion}
This paper investigated \textbf{“Has the information storage potential of INR network parameters been fully utilized?”} We demonstrated through parameter reuse that the current INR video compression network could store more information. By studying various methods of parameter reuse, we developed a scheme that enhanced rate-distortion performance by increasing network depth without adding parameters. This approach showed superior rate-distortion performance compared to other INR video compression methods on the HEVC Class B and MCL-JCV datasets. Our findings offered a new perspective for future INR video compression, suggesting that designing more efficient INR network structures with parameter reuse could significantly improve INR video compression performance.

\bibliography{aaai25}

\end{document}